\documentclass[submission,copyright,creativecommons]{eptcs}
\providecommand{\event}{HCVS 2014} 

\usepackage{amsmath,amssymb,amsfonts,amsthm}
\usepackage{graphicx}
\usepackage{listings}
\usepackage{tikz}
\usepackage{bbm}
\usepackage{proof}
\usepackage{alltt}
\usepackage{paralist}
\usetikzlibrary{arrows,automata,positioning,calc}

\tikzset{autoEdge/.style={-latex}}

\setdefaultenum{(i)}{}{}{}

\allowdisplaybreaks[1]

\newcommand{\distinct}{\ensuremath{\mathit{dist}}}

\newtheorem{definition}{Definition}
\newtheorem{example}{Example}
\newtheorem{lemma}{Lemma}
\newtheorem{theorem}{Theorem}

\newcommand{\arrow}[2]{\ifx#1\relax\else\rule{0mm}{2.3ex}\fi\smash{\xrightarrow[{\scriptstyle #2}]{{\scriptstyle #1}}}}

\newif\ifLongVersion\LongVersionfalse

\newif\ifoutline
\outlinefalse

\newcommand{\constraint}{constraint}
\newcommand{\constraintLang}{\constraint\ language}

\newcommand{\Constraint}{Constraint}
\newcommand{\ConstraintLang}{\Constraint\ language}
\newcommand{\ConstraintLangs}{\ConstraintLang s}
\newcommand{\Con}{\ensuremath{\mathit{Constr}}} 

\newcommand{\relsym}{relation symbol}
\newcommand{\relsyms}{\relsym s}

\newcommand{\unicl}{\ensuremath{\mathit{Cl}_\forall}}

\newcommand{\ClauseSet}{{\symCal HC}}

\newcommand{\symCal}{}

\lstdefinelanguage{scala}{
  alsoletter={@,=,>},
  morekeywords={def,let,abstract, case, catch, class, def, do, else, extends, false, final, finally, for, if, implicit, import, match, new, null, object, 
override, package, private, protected, requires, return, sealed, super, this, throw, trait, try, true, type, val, var, while, yield, domain, Boolean,
postcondition, precondition,invariant, constraint, assert, forAll, in, _, return, @generator, ensure, require, ensuring, given, have, =>, input, output, continue},
  sensitive=true,
  morecomment=[l]{//},
  morecomment=[s]{/*}{*/},
  morestring=[b]"
}

\newcommand{\codestyle}{\small\sffamily}

\title{Horn Clauses for Communicating Timed Systems}

\author{Hossein Hojjat
\institute{Cornell University, USA}
\and Philipp R\"ummer\quad Pavle Subotic \quad Wang Yi
\institute{Uppsala University, Sweden}
}

\def\titlerunning{Horn Clauses for Communicating Timed Systems}
\def\authorrunning{Hossein Hojjat \& Philipp R\"ummer \& Pavle Subotic \& Wang Yi }

\begin{document}
\maketitle

\begin{abstract}
Languages based on the theory of timed automata are a well established
approach for modelling and analysing real-time systems, with many
applications both in industrial and academic context. Model checking
for timed automata has been studied extensively during the last two
decades; however, even now industrial-grade model checkers are
available only for few timed automata dialects (in particular Uppaal
timed automata), exhibit limited scalability for systems with large
discrete state space, or cannot handle parametrised systems. We
explore the use of Horn constraints and off-the-shelf model checkers
for analysis of networks of timed automata.  The resulting analysis
method is fully symbolic and applicable to systems with large or
infinite discrete state space, and can be extended to include various
language features, for instance Uppaal-style communication/broadcast
channels and BIP-style interactions, and systems with infinite
parallelism. Experiments demonstrate the feasibility of the method.


\end{abstract}

\section{Introduction}

We consider the analysis of
systems with real-time aspects, a problem that is commonly addressed with the
help of \emph{timed automata} models.  By modelling systems as timed
automata, a variety of relevant properties can be analysed, including
schedulability, worst-case execution time of concurrent systems,
interference, as well as functional properties. Tools and model
checking techniques for timed automata have been studied extensively
during the last two decades, one prime example being the Uppaal
tool~\cite{uppaal}, which uses difference-bound matrices (DBMs) for the
efficient representation of time, and explicit representation of data
(discrete state). Despite many advances, scalability of tools for
analysing timed automata remains a concern, in particular for models
of industrial size.

We investigate the use of fully-symbolic model checking for the
analysis of timed systems, leveraging counterexample-guided
abstraction refinement
(CEGAR)~\cite{GrafSaidi97ConstructionAbstractStateGraphsPVS,%
  HenzingerETAL04AbstractionsProofs} to represent state space, with
Craig interpolation~\cite{craig1957linear} for the refinement step, as
well as the recently proposed framework of Horn
clauses~\cite{lopstr07,DBLP:conf/popl/GuptaPR11} as intermediate
system representation. Symbolic methods enable us to handle timed
systems that are beyond the capabilities of DBM-based model checkers,
due to the size of the discrete state space (which in realistic models
can be large, or even infinite). The flexibility of Horn constraints
makes it possible to elegantly encode language features of timed
systems that are commonly considered difficult; in particular, systems
with an unbounded (or infinite) number of processes can be handled in
much the same way as bounded systems. At the same time, Horn
constraints enable the application of various general-purpose model
checkers, for instance Z3~\cite{DBLP:conf/sat/HoderB12},
Q'ARMC~\cite{andrey-pldi}, or Eldarica~\cite{eldarica-tool}, and
streamline the engineering of analysis tools.

Contributions of the paper are:
\begin{inparaenum}
\item a uniform Horn clause encoding for systems with (finite or
  infinite) concurrency, real-time constraints, as well as
  inter-process communication using shared memory, synchronous message
  passing, and synchronisation using barriers; the encoding can be
  applied, among others, to Uppaal timed automata~\cite{uppaal} and
  BIP~\cite{DBLP:conf/atva/BensalemBSN08};
\item an experimental evaluation using a set of (well-known)
  parametric timed automata models.
\end{inparaenum}



\subsection{Related Work}

We focus on work most closely related to ours; for a general overview
of timed automata analysis, the reader is referred to surveys like
\cite{DBLP:journals/csr/WaezDR13}.

Our work is inspired by recent results on the use of \textbf{Horn
  clauses} for concurrent system analysis, in particular Owicki-Gries
and Rely-Guarantee approaches in~\cite{andrey-pldi}. We use
Owicki-Gries-style invariants in our work, but generalise the way how
invariants can relate different processes of a system (using
\emph{invariant schemata}), and include systems with infinitely many
processes and time. For parametric systems, we generate invariants
quantifying over all processes in a system; the way such invariants
are derived has similarities to \cite{DBLP:conf/sas/BjornerMR13},
where quantified invariants are inferred in the context of datatypes
like arrays. Encoding of timed automata as Horn clauses has also been
proposed in
\cite{DBLP:conf/sat/HoderB12,DBLP:journals/mics/FietzkeW12}, but only
restricted to derivation of monolithic system invariants
(non-compositional reasoning). Similarly, there is work on
non-compositional/parametric analysis of timed systems using
logic programming (CLP)
\cite{DBLP:conf/rtss/GuptaP97,DBLP:conf/rtss/JaffarSV04,DBLP:conf/lopstr/BandaG08,DBLP:journals/tplp/FioravantiPPS13}.

\textbf{$k$-Indexed invariants} were introduced in
\cite{DBLP:conf/sas/SanchezSSC12}, as an instance of the general
concept of \textbf{thread-modular model checking}
\cite{DBLP:conf/spin/FlanaganQ03}, using self-reflection to
automatically build environments of threads. We carry over the
approach to Horn clauses, and investigate its use for extensions of timed
systems.

SMT-based full model checking (\textbf{$k$-induction} and
\textbf{IC3}) for timed automata has recently been investigated in
\cite{DBLP:conf/formats/KindermannJN12}, using the region abstraction
for discretisation. In comparison, our work relies on CEGAR to handle
time and data alike, and achieves compositional and parametric
analysis via  Horn clauses.

The approach of \textbf{backward reachability} has been used for
verification of various classes of parametric systems, including timed
systems~\cite{DBLP:journals/tcs/AbdullaJ03,DBLP:conf/lics/AbdullaDM04,mcmtexperiments},
establishing decision procedures with the help of suitable syntactic
restrictions.  A detailed comparison between backward reachability for
timed systems and our approach is beyond the scope of this paper, and
is planned as future work. Since our approach naturally includes
abstraction through CEGAR, we expect better scalability for systems
with complex process-local behaviour (e.g., if individual processes
are implemented as software programs). On the other hand, backward
reachability gives rise to decision procedures for important classes
of parametric systems; it is unclear whether such results can be
carried over to our setting.



\section{Motivating Examples}

\subsection{Railway Control System\label{railway}}

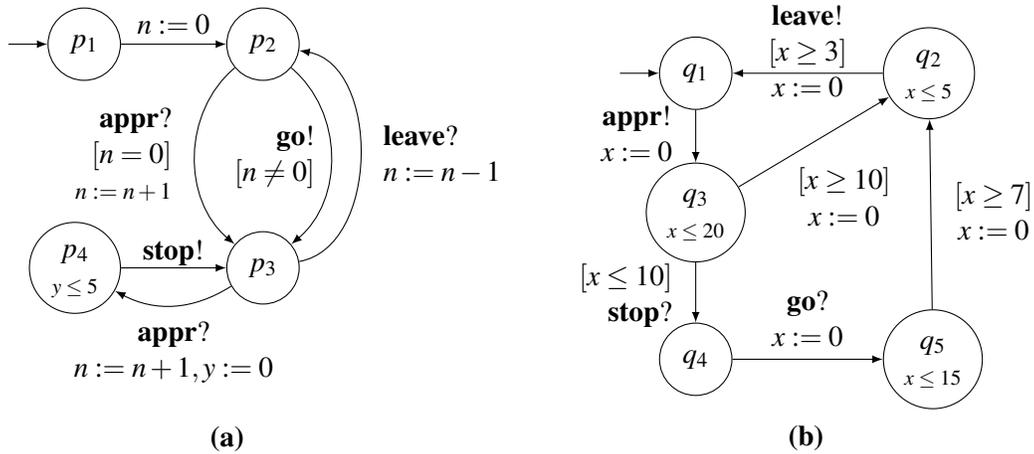
\begin{figure}[tb]
\begin{minipage}{.53\textwidth}
\centering
\mbox{}\\[1ex]
\qquad
\begin{tikzpicture}
  \tikzstyle{initial}=[initial by arrow,initial text=]
  \tikzstyle{every initial by arrow}=[-latex]

  \node[state,initial] (1) {$p_1$};
  \node[state,right=1.4cm of 1] (2) {$p_2$};
  \node[state,below =2cm of 2] (3) {$p_3$};
  \node[state,left=1.4cm of 3,inner sep=-0.5mm] (4) {
        $\begin{array}{c}
        p_4\\
        \mbox{{\scriptsize $y\le5$}}
      \end{array}$
  };

  \path[autoEdge] (1) edge node[auto] {$n:=0$}   (2);
  \path[autoEdge,bend right=80] (3) edge
                 node[right] {
             $\begin{array}{l}
                {\bf leave}?
                \\
                n:=n-1
             \end{array}$} (2);
  \path[autoEdge,bend right=50] (2) edge
                 node[left] {
             $\begin{array}{r}
                {\bf appr}?
                \\{}
                [n=0]
                \\
                \mbox{{\footnotesize $n:=n+1$}}
             \end{array}$} (3);
  \path[autoEdge,bend left=50] (2) edge
                 node[left=-0.1] {
             $\begin{array}{r}
               {\bf go}!
               \\{}
               [n\neq0]
             \end{array}$} (3);
  \path[autoEdge,bend left=30] (3) edge
                 node[below] {
             $\begin{array}{c}
                {\bf appr}?
                \\
                n:=n+1, y := 0
             \end{array}$} (4);
  \path[autoEdge] (4) edge
                 node[above=-0.1] {
                 ${\bf stop}!$} (3);

\end{tikzpicture}
\\[1ex]
{\bf (a)}
\end{minipage}
\begin{minipage}{.42\textwidth}
\centering
\begin{tikzpicture}
  \tikzstyle{initial}=[initial by arrow,initial text=]
  \tikzstyle{every initial by arrow}=[-latex]

  \node[state,initial] (1) {$q_1$};
  \node[state,right=2cm of 1,auto,inner sep=-0.5mm] (2) {
      $\begin{array}{c}
        q_2
        \\
        \mbox{{\scriptsize $x\le5$}}
      \end{array}$
  };
  \node[state,below=0.7cm of 1,inner sep=-0.5mm] (3) {
      $\begin{array}{c}
        q_3\\
        \mbox{{\scriptsize $x\le20$}}
      \end{array}$
  };
  \node[state,below=0.8cm of 3] (4) {$q_4$};
  \node[state,right=2cm of 4,inner sep=-0.5mm] (5) {
      $\begin{array}{c}
        q_5
        \\
        \mbox{{\scriptsize $x\le15$}}
       \end{array}$
  };

  \path[autoEdge] (2) edge 
                 node[above,above=-0.59] {
             $\begin{array}{c}
                {\bf leave}!
                \\
                {[}x\ge3]
                \\
                x := 0
             \end{array}$}   (1);
  \path[autoEdge] (1) edge
                 node[left] {
             $\begin{array}{c}
                {\bf appr}!
                \\
                x :=0
             \end{array}$} (3);
  \path[autoEdge] (3) edge
                 node[left,pos=0.55] {
             $\begin{array}{r}
                [x\le10]
                \\
                {\bf stop}?
             \end{array}$} (4);
  \path[autoEdge] (4) edge
                 node[above,above=-0.1] {
             $\begin{array}{c}
                {\bf go}?
                \\
                x := 0
             \end{array}$} (5);
  \path[autoEdge] (5) edge
                 node[right] {
             $\begin{array}{c}
                [x\ge7]
                \\
                x := 0
             \end{array}$} (2);
  \path[autoEdge] (3) edge
                 node[below=0.4,pos=0.7] {
             $\begin{array}{c}
                [x\ge10]
                \\
                x := 0
             \end{array}$} (2);
\end{tikzpicture}
\\
{\bf (b)}
\end{minipage}
\caption{Railway Control System~\cite{DBLP:conf/forte/YiPD94}. (a)
  Controller, (b) Train.\label{fig::train-example}}
\end{figure}

Fig.~\ref{fig::train-example} depicts a train controller system taken
from~\cite{DBLP:conf/forte/YiPD94}, consisting of a number of trains
travelling towards a critical point that can be passed by only one
train at a time, and a controller responsible for preventing
collisions. Compared to \cite{DBLP:conf/forte/YiPD94}, the model was
simplified by removing the queue to store incoming requests from the
trains; since we only focus on safety, not fairness, this queue
becomes irrelevant.
The controller randomly releases trains without considering any
specific order.

The trains communicate with the controller using binary communication
channels.  When a train approaches the critical point it informs the
controller via the channel {\bf appr}.  It then waits for $20$ time
units; if the controller does not stop the train using
{\bf stop}, it enters its crossing state $q_2$. 
As a safety property of the system we require that at any time only
one train can be in $q_2$; using Uppaal, the authors
of~\cite{DBLP:conf/forte/YiPD94} could successfully prove safety for
up to $6$ trains. In our setting, we consider the model with
\emph{infinitely} many instances of the train automaton; this subsumes
the parametric problem of showing safety for an arbitrary (finite)
number~$N$ of trains.  We show safety of the infinite model
(automatically) by computing a quantified inductive invariant of the
form $\forall \mathit{id}_1, \mathit{id}_2, \mathit{id}_3.\;
I(\mathit{ctrl}, \mathit{train}(\mathit{id}_1),
\mathit{train}(\mathit{id}_2), \mathit{train}(\mathit{id}_3))$, where
$\mathit{ctrl}$ is the state of the controller, and
$\mathit{train}(\mathit{id})$ the state of a specific
train~$\mathit{id}$; in other words, the invariant expresses a
property that holds for any triplet of trains at any time. Note that
invariants of this kind can express that at most two trains are
in $q_3$.



\subsection{RT-BIP Example}
\label{sec:bipExample}

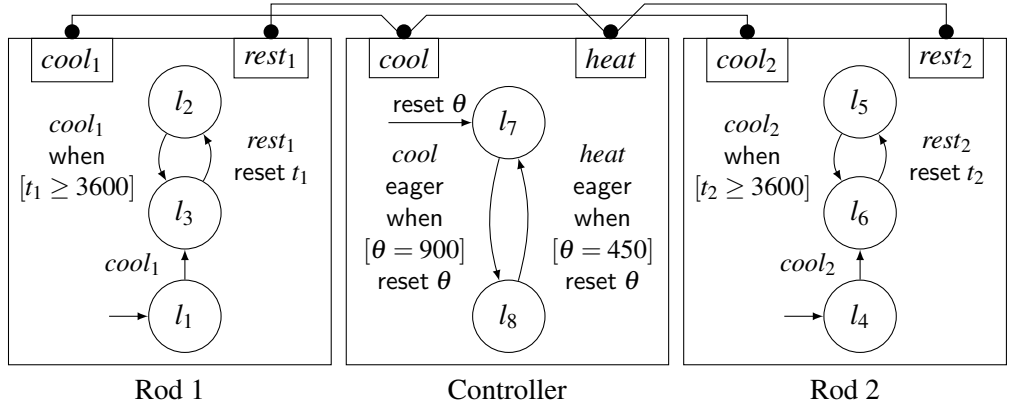
\begin{figure}[tb]
  \centering
\begin{tikzpicture}
  \tikzstyle{initial}=[initial by arrow,initial text=]
  \tikzstyle{every initial by arrow}=[-latex]

  \node[state,initial] (1) {$l_1$};
  \node[state,above=0.4cm of 1] (3) {$l_3$};
  \node[state,above=0.5cm of 3] (2) {$l_2$};

  \path[autoEdge] (1) edge node[auto] {\small
             $\begin{array}{c}
               \mathit{cool}_1
              \end{array}$}   (3);
  \path[autoEdge,bend right] (2) edge
                 node[auto,swap] {\small
             $\begin{array}{c}
                \mathit{cool}_1
                \\
                \mbox{\small\textsf{when}}
                \\{}
                \mbox{\small $[t_1 \geq 3600]$}
             \end{array}$} (3);
  \path[autoEdge,bend right] (3) edge
                 node[auto,swap] {\small
             $\begin{array}{c}
                \mathit{rest}_1
                \\
                \mbox{\small\textsf{reset}~$t_1$}
              \end{array}$} (2);

  \draw ($(1.south west)+(-2,-0.3)$)  coordinate (A1)
       rectangle ($(2.north east)+(1.6,0.5)$) coordinate (B1);
  \node[shape=rectangle,draw,anchor=north west] (cool1)
     at ($(A1 |- B1)+(0.3,0.2pt)$) {$\mathit{cool}_1$};
  \node[shape=rectangle,draw,anchor=north east] (rest1)
     at ($(B1)+(-0.3,0.2pt)$) {$\mathit{rest}_1$};


  \node[state,initial,right=8 of 1] (4) {$l_4$};
  \node[state,above=0.4cm of 4] (6) {$l_6$};
  \node[state,above=0.5cm of 6] (5) {$l_5$};

  \path[autoEdge] (4) edge node[auto] {\small
             $\begin{array}{c}
               \mathit{cool}_2
              \end{array}$}   (6);
  \path[autoEdge,bend right] (5) edge
                 node[auto,swap] {\small
             $\begin{array}{c}
                \mathit{cool}_2
                \\
                \mbox{\small\textsf{when}}
                \\{}
                \mbox{\small $[t_2 \geq 3600]$}
             \end{array}$} (6);
  \path[autoEdge,bend right] (6) edge
                 node[auto,swap] {\small
             $\begin{array}{c}
                \mathit{rest}_2
                \\
                \mbox{\small\textsf{reset}~$t_2$}
              \end{array}$} (5);

  \draw ($(4.south west)+(-2,-0.3)$)  coordinate (A2)
       rectangle ($(5.north east)+(1.6,0.5)$) coordinate (B2);
  \node[shape=rectangle,draw,anchor=north west] (cool2)
     at ($(A2 |- B2)+(0.3,0.2pt)$) {$\mathit{cool}_2$};
  \node[shape=rectangle,draw,anchor=north east] (rest2)
     at ($(B2)+(-0.3,0.2pt)$) {$\mathit{rest}_2$};


  \node[state,right=3.33 of 1,yshift=0cm] (8) {$l_8$};
  \node[state,above=1.6cm of 8] (7) {$l_7$};

  \path[autoEdge] ($(7)+(-1.6,0)$) edge
                 node[auto] {\small\textsf{reset}~$\theta$} (7);

  \path[autoEdge,bend right=15] (8) edge
                 node[auto,swap] {\small
             $\begin{array}{c}
                \mathit{heat}
                \\
                \mbox{\small\textsf{eager}}
                \\
                \mbox{\small\textsf{when}}
                \\
                \mbox{\small $[\theta = 450]$}
                \\
                \mbox{\small\textsf{reset}~$\theta$}
             \end{array}$} (7);
  \path[autoEdge,bend right=15] (7) edge
                 node[auto,swap] {\small
             $\begin{array}{c}
                \mathit{cool}
                \\
                \mbox{\small\textsf{eager}}
                \\
                \mbox{\small\textsf{when}}
                \\
                \mbox{\small $[\theta = 900]$}
                \\
                \mbox{\small\textsf{reset}~$\theta$}
              \end{array}$} (8);

  \draw ($(A1 -| B1)+(0.2,0)$) coordinate (A3)
       rectangle ($(A2 |- B2)+(-0.2,0)$) coordinate (B3);
  \node[shape=rectangle,draw,anchor=north west] (cool)
     at ($(A3 |- B3)+(0.3,0.2pt)$) {$\mathit{cool}$};
  \node[shape=rectangle,draw,anchor=north east] (heat)
     at ($(B3)+(-0.3,0.2pt)$) {$\mathit{heat}$};


  \draw (cool.north) -- +(-0.3,0.3) -| (cool1.north);
  \draw (cool.north) -- +(0.3,0.3) -| (cool2.north);
  \draw (heat.north) -- +(-0.45,0.45) -| (rest1.north);
  \draw (heat.north) -- +(0.45,0.45) -| (rest2.north);

  \fill ($(cool.north)+(0,0.08)$) circle (0.1);
  \fill ($(heat.north)+(0,0.08)$) circle (0.1);
  \fill ($(cool1.north)+(0,0.08)$) circle (0.1);
  \fill ($(rest1.north)+(0,0.08)$) circle (0.1);
  \fill ($(cool2.north)+(0,0.08)$) circle (0.1);
  \fill ($(rest2.north)+(0,0.08)$) circle (0.1);

  \node[anchor=north,yshift=-0.6mm] at ($(A1)!0.5!(A1 -| B1)$) {Rod 1};
  \node[anchor=north,yshift=-0.6mm] at ($(A2)!0.5!(A2 -| B2)$) {Rod 2};
  \node[anchor=north,yshift=-0.6mm] at ($(A3)!0.5!(A3 -| B3)$) {Controller};
\end{tikzpicture}
\caption{Temperature Control System~\cite{DBLP:conf/atva/BensalemBSN08}\label{fig::bip-example}}
\end{figure}

Figure~\ref{fig::bip-example} shows a temperature control system
modeled in the component coordination language
BIP~\cite{DBLP:conf/atva/BensalemBSN08}.  The Controller component is
responsible for keeping the value of $\theta$ between the values $450$
and $900$.  Whenever the value of $\theta$ reaches the upper bound of
$900$ the Controller sends a cooling command to the Rod 1 and Rod 2
components using its {\it cool} port.  In the lower bound of $450$ the
Controller resets the Rods.  The Rod components can accept a {\it
  cool} command again only if $3600$ time units have elapsed.
Compared to \cite{DBLP:conf/atva/BensalemBSN08}, we use the RT-BIP
dialect and model both physical time~$t_1, t_2$ and
temperature~$\theta$ using clocks.
 Note that in this model all
the communications are in the form of Rendezvous so the priority layer
of the BIP glue is essentially empty.  The required safety property of
the system is to ensure deadlock freedom.  Deadlock happens when the
value of $\theta$ reaches $900$ but there was not sufficient time
($3600$ time units) for the rods to be engaged again.

BIP semantics requires that all processes run to an interaction point,
before interaction takes place. We model this using a basic form of
synchronization barrier (Section~\ref{sec:bip}), together with a
global variable~$\mathit{iact}$ to choose between interactions.  Among
a set of parallel processes that share a barrier, whenever a process
reaches the barrier it stops until all the other processes reach the
barrier.  Verification shows that the model has a deadlock; the
heating period of the controller is faster than the required delay
time of the rods.





\section{Preliminaries}

\paragraph{\ConstraintLangs.}
Throughout this paper, we assume that a first-order vocabulary of
\emph{interpreted symbols} has been fixed, consisting of a
set~$\Sigma_f$ of fixed-arity function symbols, and a set~$\Sigma_p$
of fixed-arity predicate symbols. Interpretation of $\Sigma_f$ and
$\Sigma_p$ is determined by a fixed structure~$(U, I)$, consisting of
a non-empty universe~$U$, and a mapping~$I$ that assigns to each
function in $\Sigma_f$ a set-theoretic function over $U$, and to each
predicate in $\Sigma_p$ a set-theoretic relation over $U$. As a
convention, we assume the presence of an equation symbol~``$=$'' in
$\Sigma_p$, with the usual interpretation.  Given a set~$X$ of
variables, a \emph{\constraintLang} is a
set~\Con\ of first-order formulae over $\Sigma_f, \Sigma_p, X$.  For example,
the language of quantifier-free Presburger arithmetic (mainly used in
this paper) has $\Sigma_f = \{+, -, 0, 1, 2, \ldots\}$ and $\Sigma_p =
\{=, \leq, |\}$, with the usual semantics.
We write $ \distinct(x_1, \ldots, x_n) ~\equiv~ \big( \forall i, j \in
\{1, \ldots, n\}.\; (i = j \vee x_i \not= x_j) \big)$ to state that
the values $x_1, \ldots, x_n$ are pairwise distinct.

\paragraph{Horn Clauses.}
We consider a set~$\symCal R$ of uninterpreted fixed-arity relation 
symbols. 
A \emph{Horn clause} is a formula~$H \leftarrow C \wedge B_1 \wedge
\cdots \wedge B_n$ where
\begin{itemize}
\item $C$ is a constraint over $\Sigma_f, \Sigma_p, X$;
\item each $B_i$ is an application~$p(t_1, \ldots, t_k)$ of a relation 
symbol $p \in \symCal R$ to first-order terms over $\Sigma_f, X$;
\item $H$ is similarly either an
  application~$p(t_1, \ldots, t_k)$ of $p \in \symCal R$ to first-order terms,
  or $\mathit{false}$.
\end{itemize}

$H$ is called the \emph{head} of the clause, $C \wedge B_1 \wedge
\cdots \wedge B_n$ the \emph{body.}  In case $C = \mathit{true}$, we
usually leave out $C$ and just write $H \leftarrow B_1 \wedge \cdots
\wedge B_n$.  First-order variables in a clause are 
implicitly universally quantified; relation symbols represent
set-theoretic relations over the universe~$U$ of a structure~$(U,
I)\in \symCal S$.  Notions like (un)satisfiability and entailment
generalise to formulae with relation symbols.

\begin{definition}[Solvability]
  Let $\ClauseSet$ be a set of Horn clauses over \relsyms\ $\symCal
  R$.  $\ClauseSet$ is called \emph{(semantically) solvable} (in the
  structure~$(U, I)$) if there is an interpretation of the
  \relsyms~$\symCal R$ as set-theoretic relations such that the universal
  closure~$\unicl(h)$ of every clause~$h \in \ClauseSet$ holds in
  $(U,I)$; in other words, if the structure~$(U, I)$ can be extended to a
  model of the clauses~$\ClauseSet$.
\end{definition}

We can practically check solvability of sets of Horn clauses by means
of \emph{predicate
  abstraction}~\cite{andrey-pldi,disjunctive-interpolant}, using tools
like Z3~\cite{DBLP:conf/sat/HoderB12}, Q'ARMC~\cite{andrey-pldi}, or
Eldarica~\cite{eldarica-tool}.



\section{Basic Encoding of Concurrent Systems}

\subsection{Semantics of Concurrent Systems}
\label{sec:semantics}

We work in the context of a simple, but expressive system model with
finitely or infinitely many processes executing concurrently in
interleaving fashion; in subsequent sections, further features like
communication will be added. Each process has its own local state
(taken from a possibly infinite state space), and in addition the
system as a whole also has a (possibly infinite) global state that can
be accessed by all processes. We use the following notation:
\begin{itemize}
\item $G$ is a non-empty set representing the global state space.
\item $P$ is a non-empty index set representing processes in the
  system.
\item The non-empty set $L_p$ represents the local state space of a
  process~$p \in P$.
\item $\mathit{Init}_p \subseteq G \times L_p$ is the set of initial
  states of a process~$p \in P$.
\item $(g, l) \stackrel{p}{\to} (g', l')$ is the transition relation
  of a process~$p \in P$, with global states~$g, g' \in G$ and local
  states~$l, l' \in L_p$.
\end{itemize}
Given a set of processes defined in this manner, we can derive a
system by means of parallel composition:
\begin{itemize}
\item $S = G \times \prod_{p \in P} L_p$ is the system state
  space. Given a system state~$s = (g, \bar l) \in S$, we write $\bar
  l[p] \in L_p$ for the local state belonging to process~$p \in P$.
\item $S_0 = \{ (g, \bar l) \mid \forall p \in P.\; (g, \bar l[p]) \in
  \mathit{Init}_p\} \subseteq S$ is the set of initial system states.
\item The transition relation of the system as a whole is defined by:
  \begin{equation*}
    \infer
    {(g, \bar l) \to (g', \bar l[p/l'])}
    {p \in P &\quad
      (g, \bar l[p]) \stackrel{p}{\to} (g', l')}
  \end{equation*}
  We write $\bar l[p/l'] \in \prod_{p \in P} L_p$ for the state vector
  obtained by updating the component belonging to process~$p \in P$ to
  $l' \in L_p$.
\end{itemize}

\paragraph{Safety.}
\hspace{-1em}
We are interested in checking \emph{safety properties} of systems as
defined above. We define (un)safety in the style of coverability, by
specifying a vector~$(\langle p_1, E_1\rangle, \ldots, \langle p_m,
E_m\rangle)$ of process-state-pairs, where the $p_i \in P$ are
pairwise distinct, and $E_i \subseteq G \times L_{p_i}$ for each $i
\in \{1, \ldots, m\}$.  System error states are:
\begin{equation*}
  \mathit{Err} = \big\{ (g, \bar l) \in S \mid
  \forall i.\; (g, \bar l[p_i]) \in E_i \big\}
\end{equation*}
Intuitively, a system state is erroneous if it contains $m$ (pairwise
distinct) processes whose state is in $E_1, \ldots, E_m$,
respectively.  Error properties like occurrence of local runtime
exceptions or violation of mutual exclusion can be expressed using
this concept of error; for instance, the property that the processes~$p_1,
p_2$ cannot reside in some state~$(g, l)$ together
is captured by $(\langle p_1, \{(g, l)\}\rangle, \langle p_2, \{(g,
l)\}\rangle)$.  A system is \emph{safe} if there is no sequence~$s_0 \to
s_1 \to \ldots \to s_n$ of transitions such that $s_0 \in S_0$ and
$s_n \in
\mathit{Err}$.


\subsection{Encoding Safety of Finite Systems}
\label{sec:finite}

To check that a system is safe it is sufficient to find an
\emph{inductive invariant,} which is a set~$\mathit{Inv} \subseteq S$
of states with the properties
\begin{inparaenum}
\item Initiation: $S_0 \subseteq \mathit{Inv}$;
\item Consecution: for all $s \to t$ with $s \in \mathit{Inv}$, also
  $t \in \mathit{Inv}$; and
\item Safety: $\mathit{Inv} \cap \mathit{Err} = \emptyset$.
\end{inparaenum}

The following sections define methods to derive inductive invariants
with the help of Horn constraints.  We first concentrate on the case
of a finite set~$P = \{1, 2, \ldots, n\}$ of processes, and show how
an encoding in the spirit of Owicki-Gries~\cite{owicki-gries} can be
done with the help of Horn constraints. In comparison to earlier
work~\cite{andrey-pldi}, inductive invariants can be defined to cover
individual processes, as well as sets of processes, in order to handle
required relational information (inspired by the concept of
\emph{$k$-indexed invariants}~\cite{DBLP:conf/sas/SanchezSSC12}). We
define this concept formally with the help of \emph{invariant
  schemata.}

Recall that the component-wise order~$<$ on the set~$\mathbbm{N}^n$ is
a well-founded partial order. An \emph{antichain} is a set~$A
\subseteq \mathbbm{N}^n$ whose elements are pairwise $<$-incomparable;
as a consequence of Dickson's lemma, antichains over $\mathbbm{N}^n$
are finite.  An \emph{invariant schema} for the processes $P = \{1, 2,
\ldots, n\}$ is an antichain $A \subseteq \{0, 1\}^n \subseteq
\mathbbm{N}^n$. Intuitively, every vector in $A$ represents an
invariant to be inferred; entries with value~$1$ in the vector
indicate processes included in the invariant, while processes with
entry $0$ are not visible (entries $\mbox{}>1$ are
relevant in Sect.~\ref{sec:hetero}).

\begin{example}
  Consider the RT-BIP model in Sect.~\ref{sec:bipExample}, with
  processes~$P = \{1, 2, 3\}$ ($1 \cong \text{Rod 1}$, $2 \cong
  \text{Controller}$, $3 \cong \text{Rod 2}$). Schema $A_1 = \{(1, 0,
  0), (0, 1, 0), (0, 0, 1)\}$ leads to fully modular safety analysis
  with three local invariants, each of which refers to exactly one
  process.  $A_2 = \{(1, 1, 0), (0, 1, 1)\}$ introduces two
  invariants, each relating one cooling rod with the controller. The
  strongest invariant schema, $A_3 = \{(1, 1, 1)\}$ corresponds to
  analysis with a single monolithic invariant.
\end{example}

To define the system invariant specified by a schema~$A$, we assume
that $\{R_{\bar a} \mid \bar a \in A\}$ is a set of relation
variables, later to be used as vocabulary for Horn
constraints. Further, given a local state vector~$\bar l \in \prod_{p
  \in R} L_p$ and $\bar a \in A$, we write $\bar l[\bar a]$ for the
vector $(\bar l[i_1], \bar l[i_2], \ldots, \bar l[i_k])$, where $i_1 <
i_2 < \cdots < i_k$ are the indexes of non-zero entries in $\bar a =
(a_1, \ldots, a_n)$ (i.e., $\{i_1, \ldots, i_k\} = \{i \in \{1, 2,
\ldots, n\} \mid a_i > 0\}$).  The system invariant is then defined as
the conjunction of the individual relation symbols~$R_{\bar a}$,
applied to global and selected local states:
\begin{equation}
  \label{eq:finiteInv}
  \mathit{Inv}(g, \bar l) ~=~
  \bigwedge_{\bar a \in A} R_{\bar a}(g, \bar l[\bar a])
\end{equation}

\def\q#1{\mathsf{#1}}

\begin{figure}[tb]
  \begin{align}
    \label{eq:finite1}
    \Big\{~
    &
    R_{\bar a}(\q g, \q l_1, \ldots, \q l_k) ~\leftarrow~
    \mathit{Init}_{i_1}(\q g, \q l_1)
    \wedge\cdots\wedge \mathit{Init}_{i_k}(\q g, \q l_k)
    ~\Big\}_{\bar a \in A}
    \\
    \label{eq:finite2}
    \Big\{~
    &
      R_{\bar a}(\q g', \bar{\q{l}}[p/\q l'][\bar a])
      ~\leftarrow~
      \big((\q g, \bar{\q{l}}[p]) \stackrel{p}{\to}
      (\q g', \q l')\big)\wedge
      R_{\bar a}(\q g, \bar{\q{l}}[\bar a])\wedge
      \mathit{Ctxt}(\{p\}, \q g, \bar{\q{l}})
    ~\Big\}_{\substack{p = 1, \ldots, n \\ \bar a \in A}}
    \\
    \label{eq:finite3}
    &
    \mathit{false} ~\leftarrow~
    \Big(
    \bigwedge_{j=1, \ldots, m}\!\!\!
    (\q g, \bar{\q{l}}[p_j]) \in E_j \Big) \wedge
    \mathit{Ctxt}(\{p_1, \ldots, p_m\}, \q g, \bar{\q{l}})
  \end{align}

  \caption{Horn constraints encoding a finite system. In
    \eqref{eq:finite1}, the numbers $i_1, \ldots, i_k$ are the indexes of non-zero
    entries in $\bar a$.  Symbols in \textsf{sans serif} are
    implicitly universally quantified variables.
  }
  \label{fig:finiteHorn}
\end{figure}

Concrete solutions for the variables $\{R_{\bar a} \mid \bar a \in
A\}$, subject to the conditions \emph{Initiation, Consecution,} and
\emph{Safety} given in the beginning of this section, can be computed
by means of an encoding as Horn clauses. For this purpose, we assume
that a system can be represented within some constraint language, for
instance within Presburger arithmetic: the sets $\mathit{Init}_p$, the
transition relation~$\smash{s \stackrel{p}{\to} t}$, as well as the error
specification~$(\langle p_1, E_1\rangle, \ldots, \langle p_m,
E_m\rangle)$ are encoded as constraints in this language. Horn clauses
can then be formulated as shown in Fig.~\ref{fig:finiteHorn}.
Clause~\eqref{eq:finite1} represents initiation, for each of the
variables~$R_{\bar a}$; \eqref{eq:finite2} is consecution, and expresses
that every relation~$R_{\bar a}$ is preserved by transitions of
any process~$p$, and
\eqref{eq:finite3} encodes unreachability of error states.

In \eqref{eq:finite2}, \eqref{eq:finite3}, we refer to a \emph{context
  invariant}~$\mathit{Ctxt}(\{p_1, \ldots, p_k\}, g, \bar l)$, which
includes those literals from $\mathit{Inv}(g, \bar l)$ relevant for
the processes~$p_1, \ldots, p_k$:
\begin{equation*}
  \mathit{Ctxt}(Q, g, \bar l) ~=~
  \bigwedge
  \big\{
  R_{\bar c}(g, \bar l[\bar c])
  \mid
  \bar c \in A \text{~and~} \exists q \in Q.\; \bar c[q] > 0
  \big\}
\end{equation*}
However, note that different choices can be made concerning invariants
$R_{\bar a}$ to be mentioned in the body of \eqref{eq:finite2},
\eqref{eq:finite3}; it is in principle possible to add arbitrary
literals from $\mathit{Inv}(g, \bar l)$. Adding more literals results
in constraints that are weaker, and potentially easier to satisfy, but
can also introduce irrelevant information.

\begin{lemma}[Soundness]
  If the constraints in Fig.~\ref{fig:finiteHorn} are
  solvable for some invariant schema~$A$, then the analysed system is safe.
\end{lemma}


\begin{lemma}[Completeness]
  \label{lem:completeness}
  If a system is safe, then there exists an invariant schema~$A$ such that
  the constraints in Fig.~\ref{fig:finiteHorn} are (semantically)
  solvable.
\end{lemma}


It is important to note that Lem.~\ref{lem:completeness} talks about
\emph{semantic solvability.} Despite existence of such a
model-theoretic solution, in general there is no guarantee that a
\emph{symbolic solution} exists that can be expressed as a formula of
the chosen constraint language. However, such guarantees can be
derived for individual classes of systems, for instance for the case
that the considered system is a network of timed automata (and a
suitable constraint language like linear arithmetic).


\subsection{Counterexample-Guided Refinement of Invariant Schemata}
\label{section:cegris}
The question remains how it is practically possible to find invariant
schemata that are sufficient to find inductive invariants.  This
aspect can be addressed via a counterexample-guided refinement
algorithm.
Initially, verification is attempted using the weakest invariant
schema, $A_0 = \{(1, 0, 0, \ldots), (0, 1, 0, \ldots), \ldots\}$.  If
verification is impossible, a Horn solver will produce a concrete
counterexample to solvability of the generated Horn constraints.  It
can then be checked whether this counterexample points to genuine
unsafety of the system, or just witnesses insufficiency of the
invariant schema.  In the latter case, a stronger invariant schema can
be chosen, and verification is reattempted.



\section{Safety for Unbounded Systems}


\subsection{Encoding of Unbounded Homogeneous Systems}
\label{sec:homo}

We now relax the restriction that the process index set~$P$ of a
system is finite, and also consider an infinite number of
processes. Since our definition of safety only considers finite paths
into potential error states, this represents the case of systems with
an unbounded number of (active) threads. Showing safety for a system with infinitely many
processes raises the challenge of reasoning about a state vector with
infinitely many entries. This can be addressed by exploiting the
symmetry of the system, by deriving a single parametric invariant that
is inductive for each process; the corresponding system invariant
universally quantifies over all processes.  Necessary relational
information pertaining to multiple processes can be captured with the
help of \emph{$k$-indexed
  invariants}~\cite{DBLP:conf/sas/SanchezSSC12}. The correctness of
parametric invariants can be encoded as a finite set of Horn
constraints, which again yields an effective method to derive such
invariants automatically.

Initially we restrict attention to \emph{homogeneous} systems, in
which all processes share the same initial states and transition
relation; however, each process has access to its process id (as a
natural number), and can adapt its behaviour with respect to the
id.\footnote{By exploiting the fact that the id can be accessed, in
  fact the model in this section is as expressive as the
  (syntactically richer) one in Sect.~\ref{sec:hetero}.}  We assume
that $P = \mathbbm{N}$, $\mathit{Init}_p = \mathit{Init}$, and $L_p =
L$ for all processes~$p \in P$.  Then, for any number~$k \in
\mathbbm{N}_{> 0}$, and given a fresh relation variable~$R$, a
$k$-indexed invariant has the shape:
\begin{equation}
  \label{eq:relpattern}
  \hspace*{-1ex}
  \mathit{Inv}(g, \bar l) =
  \forall p_1, \ldots, p_k \in \mathbbm{N}.\;
  \big(
  \distinct(p_1, \ldots, p_k) \to
  R(g, p_1, \bar l[p_1], \ldots,
  p_k, \bar l[p_k])
  \big)
\end{equation}
$R$ represents a formula that can talk about the global state~$g$, as
well as about $k$ pairs~$(p_i, \bar l[p_i])$ of
(pairwise distinct) process identifiers and local process states. $R$
can therefore express which combinations of states of multiple
processes can occur simultaneously, and encode properties like mutual
exclusion (at most one process can be in some state at a
time). For $k = 1$, the invariants reduce to Owicki-Gries-style
invariants (for infinitely many processes).

\begin{figure}[tb]
  \begin{align}
    \label{eq:homo0}
    \Big\{~
    &
        R(\q g, \q p_{\sigma(1)}, \q l_{\sigma(1)}, \ldots, \q p_{\sigma(k)}, \q l_{\sigma(k)})
        ~\leftarrow~
        \distinct(\q p_1, \ldots, \q p_k) \wedge
        R(\q g, \q p_1, \q l_1, \ldots, \q p_k, \q l_k)
      ~\Big\}_{\sigma \in S_k}
    \\
    \label{eq:homo1}
    &
    R(\q g, \q p_1, \q l_1, \ldots, \q p_k, \q l_k) ~\leftarrow~
    \distinct(\q p_1, \ldots, \q p_k) \wedge
    \mathit{Init}(\q g, \q l_1)
    \wedge\cdots\wedge \mathit{Init}(\q g, \q l_k)
    \\
    \label{eq:homo2}
    &
      R(\q g', \q p_1, \q l'_1, \ldots, \q p_k, \q l_k)
      ~\leftarrow~
      \distinct(\q p_1, \ldots, \q p_k) \wedge
      \big((\q g, \q l_1) \stackrel{\q p_1}{\to}
      (\q g', \q l'_1)\big)\wedge
      R(\q g, \q p_1, \q l_1, \ldots, \q p_k, \q l_k)
    \\
    \label{eq:homo2b}
    &
      R(\q g', \q p_1, \q l_1, \ldots, \q p_k, \q l_k)
      ~\leftarrow~
      \distinct(\q p_0, \q p_1, \ldots, \q p_k) \wedge
      \big((\q g, \q l_0) \stackrel{\q p_0}{\to}
      (\q g', \q l'_0)\big)\wedge
      \mathit{RConj}(0, \ldots, k)
    \\
    \label{eq:homo3}
    &
    \mathit{false} ~\leftarrow~
    \distinct(\q p_1, \ldots, \q p_r) \wedge
    \Big(
    \bigwedge_{j=1, \ldots, m}\!\!\!
    (\q p_j = p_j \wedge
    (\q g, \q l_j) \in E_j) \Big) \wedge
    \mathit{RConj}(1, \ldots, r)
  \end{align}

  \caption{Horn constraints encoding a homogeneous infinite system
    with the help of a $k$-indexed invariant.  $S_k$ is the symmetric
    group on $\{1, \ldots, k\}$, i.e., the group of all permutations
    of $k$ numbers; as an optimisation, any generating subset of
    $S_k$, for instance transpositions, can be used instead of
    $S_k$. In \eqref{eq:homo3}, we define $r = \max\{m, k\}$.}
  \label{fig:homogeneousHorn}
\end{figure}

 Fig.~\ref{fig:homogeneousHorn} gives the Horn
clauses encoding the assumed properties of $\mathit{Inv}(g, \bar l)$
for a given~$k$.  Since $k$-indexed invariants quantify over all
permutations of $k$ processes, it can be assumed that $R$ is
symmetric, which is captured by \eqref{eq:homo0}. Initiation is
encoded in \eqref{eq:homo1}. Consecution is split into two cases:
\eqref{eq:homo2} covers the situation that $\q p \in \{\q p_1, \ldots,
\q p_k\}$ makes a transition, and \eqref{eq:homo2b} for transitions
due to some process $\q p_0 \not\in \{\q p_1, \ldots, \q p_k\}$. In
\eqref{eq:homo2}, due to symmetry of $R$, it can be assumed that $\q p
= \q p_1$. Unreachability of errors~$(\langle p_1, E_1\rangle, \ldots,
\langle p_m, E_m\rangle)$ is specified by \eqref{eq:homo3}.

As shorthand notation in \eqref{eq:homo2b}, \eqref{eq:homo3}, for
numbers~$a, b \in \mathbbm{N}$ with $a \leq b$ the expression
$\mathit{RConj}(a, \ldots, b)$ represents the conjunction of all
$R$-instances for process ids in the range $a, \ldots, b$ (as in
Sect.~\ref{sec:finite}, it is possible to include further literals
from $\mathit{Inv}(g, \bar l)$ in the body of
\eqref{eq:homo2}--\eqref{eq:homo3}, resulting in weaker constraints):
\begin{equation*}
  \mathit{RConj}(a, \ldots, b) ~~=~~
  \bigwedge_{\substack{i_1, \ldots, i_k \in \{a, \ldots, b\}\\
      i_1 < i_2 < \cdots < i_k}}
  R(\q g, \q p_{i_1}, \q l_{i_1}, \ldots, \q p_{i_k}, \q l_{i_k})
\end{equation*}

\begin{theorem}[Expressiveness]
  \label{thm:expressiveness}
  \begin{inparaenum}
  \item If the constraints in Fig.~\ref{fig:homogeneousHorn} are
    satisfiable for a given $k$, then they are also satisfiable for
    any $k' > k$ (for the same system).
  \item If $k' > k > 0$, then there are systems that can be verified
    with $k'$-indexed invariants, but not with $k$-indexed invariants.
  \end{inparaenum}
\end{theorem}


\subsection{Encoding of Unbounded Heterogeneous Systems}
\label{sec:hetero}

The encodings of Sect.~\ref{sec:finite} and \ref{sec:homo} can be
combined, to analyse systems that contain $n$ different types of
processes, each of which can either be a \emph{singleton} process, or
a process that is \emph{infinitely replicated.} Compared to
Sect.~\ref{sec:homo}, process types enable more fine-grained use of
$k$-indexed invariants, since it is now possible to specify which
processes are considered with which arity in an invariant.

More formally, we now use the process index set $P = \bigcup_{i
  = 1, \ldots n} (\{i\} \times P_i)$, where $P_i$ is either $\{0\}$
(singleton case) or $\mathbbm{N}$ (replicated case).  \emph{Invariant
  schemata} from Sect.~\ref{sec:finite} generalise to unbounded
heterogeneous systems, and are now antichains~$A$ of the partially
ordered set $\prod_{i = 1, \ldots n} (\{0, 1\} \cup P_i) \subseteq
\mathbbm{N}^n$. This means that an invariant can refer to at most one
instance of a singleton process, but to multiple instances of
replicated processes.  As in Sect.~\ref{sec:finite}, we use a set
$\{R_{\bar a} \mid \bar a \in A\}$ of relation variables, and define
the system invariant as a conjunction of individual invariants, each
of which now quantifies over ids of processes. Namely, for a
vector~$\bar a = (a_1, \ldots, a_n)$ and process type~$i \in \{1,
\ldots, n\}$, $a_i$ distinct processes~$p_1^i, \ldots, p_{a_i}^i \in
P_i$ are considered:\footnote{Note that if $i$ is a singleton process,
  there is only a single process id ($P_i = \{0\}$), so that the
  corresponding argument of $R_{\bar a}$ could be left out.}
\begin{equation*}
  \mathit{Inv}(g, \bar l) ~=
  \bigwedge_{\substack{\bar a \in A \\ \bar a = (a_1, \ldots, a_n)}}
  \hspace*{-1ex}
  \raisebox{-1ex}{$
  \begin{array}{l}
    \forall p_1^1, \ldots, p_{a_1}^1 \in P_1. \ldots
    \forall p_1^n, \ldots, p_{a_n}^n \in P_n.\\[0.5ex]
    \quad
    \big(\distinct(p_1^1, \ldots, p_{a_1}^1) \wedge \cdots \wedge
    \distinct(p_1^n, \ldots, p_{a_n}^n)\\[0.5ex]
    \qquad\to R_{\bar a}(g, p_1^1, \bar l[(1, p_1^1)], p_2^1, \bar l[(1, p_2^1)],
    \ldots, p_{a_n}^n, \bar l[(n, p_{a_n}^n)])
    \big)
  \end{array}$}
\end{equation*}
It is then possible to formulate Horn constraints about the required
properties of the invariants. The Horn clauses combine features of
those in Fig.~\ref{fig:finiteHorn} and \ref{fig:homogeneousHorn}, but
are left out from this paper due to the notational
complexity.

\begin{example}
  Consider the railway control system in
  Fig.~\ref{fig::train-example}, which consists of a singleton
  process~$P_1 = \{0\}$, the controller, and an infinitely replicated
  process~$P_2 = \mathbbm{N}$, the trains. The system can be verified
  with the schema~$A = \{(1, 3)\}$; this means, an inductive invariant
  is derived that relates the controller with a triplet of (arbitrary,
  but distinct) trains.
\end{example}




\section{Encoding of Physical Time}

We now describe how our model of execution, and the encoding as Horn
constraints, can be extended to take physical time into account. In
this and the following sections we focus on the Horn encoding of
unbounded homogeneous systems (Sect.~\ref{sec:homo}), but stress that
the same extensions are possible for heterogeneous systems
(Sect.~\ref{sec:hetero}) and finite systems (Sect.~\ref{sec:finite}).

In our system model, time is represented as a component of the global
state~$g \in G$. As a convention, we write~$g[C]$ to access the
current time, and $g' = g[C/C']$ to update time to a new value~$C' \in
\mathit{Time}$, where $\mathit{Time} = \mathbbm{Q}$ for a dense model
of time, and $\mathit{Time} = \mathbbm{Z}$ for discrete time. Time
elapse is represented by an additional rule, augmenting the transition
relation as defined in Sect.~\ref{sec:semantics}:
\begin{equation*}
  \infer[\textsf{time-elapse}]
  {(g, \bar l) \to (g[C/C'], \bar l)}
  {C' \in \mathit{Time}\quad C' \geq g[C]\quad
  \forall p \in P.\; (g[C/C'], \bar l[p]) \in \mathit{TInv}_p}
\end{equation*}
The premises state that time can only develop
monotonically, and only as long as the \emph{time
  invariant}~$\mathit{TInv}_p \subseteq G \times L_p$ of all
processes~$p \in P$ is satisfied. We make the assumption
that $\mathit{TInv}_p$ is convex with respect to time, i.e.,
$(g[C/C_1], l) \in \mathit{TInv}_p$ and $(g[C/C_2], l) \in
\mathit{TInv}_p$ imply $(g[C/C_3], l) \in \mathit{TInv}_p$ for all
$C_1 \leq C_3 \leq C_2$.

Concepts like clocks or stopwatches can easily be represented by
defining local process transitions. For instance, a clock is realised
by means of a $\mathit{Time}$-valued variable~$x$; resetting the clock
is translated to the assignment~$x := g[C]$, so that the value of the
clock at any point is $g[C] - x$.

When the model of time is dense it is still possible to retain the
global variable $C$ in the integer domain.  Considering the value of
time to be a fractional number, the variable $C$ can store the
numerator, and a new global variable~$U$ is added to the system to
represent the denominator. The variable~$U$ is initialised with an
arbitrary positive value at system start, but can then stay constant
throughout the system execution, whereas the numerator $C$ is
incremented by time elapse transitions.  The encoding of rationals
using numerator and denominator faithfully represents dense time, but
makes it possible to keep all variables integer-valued.  In practice
this is helpful when no rational solver is available.

\paragraph{Horn constraints.} On the level of inductive invariants,
time just requires to add one further clause to the constraints in
Fig.~\ref{fig:homogeneousHorn}; as before, this necessitates
sets~$\mathit{TInv}_p$ that can be represented in the constraint
language of the clauses.

\begin{equation}
  R(\q g[C/\q C'], \q p_1, \q l_1, \ldots, \q p_k, \q l_k)
  ~~\leftarrow~~
  \begin{array}{@{}l@{}}
      \distinct(\q p_1, \ldots, \q p_k) \wedge
      (\q C' \geq \q g[C]) \wedge
      R(\q g, \q p_1, \q l_1, \ldots, \q p_k, \q l_k) \wedge\mbox{}
      \\
      (\q g[C/\q C'], \q l_1) \in \mathit{TInv}_{\q p_1} \wedge
      \cdots \wedge
      (\q g[C/\q C'], \q l_k) \in \mathit{TInv}_{\q p_k}
    \end{array}
\end{equation}




\section{Communication and Synchronisation}

At this point, our model of execution supports communication between
processes via the global state of a system (shared variables). To
naturally represent timed automata models and message passing
communication, it is appropriate to introduce further communication
primitives, together with their encoding as Horn constraints, which is
done in the next sections.

\subsection{Uppaal-style Binary Communication Channels}
\label{sec:uppaalComm}

Binary communication channels in Uppaal implement a simple form of
synchronisation between pairs of processes (rendezvous). We assume
that $\mathit{Ch}$ is a finite set of channel identifiers. In addition
to local transitions~$\smash{(g, l) \stackrel{p}{\to} (g', l')}$ of a
process~$p \in P$ (as in Sect.~\ref{sec:semantics}), we then also
consider \emph{send} transitions $\smash{(g, l) \stackrel{p,\,
    a!}{\longrightarrow} (g', l')}$ and \emph{receive} transitions
$\smash{(g, l) \stackrel{p,\, a?}{\longrightarrow} (g', l')}$ for any
communication channel~$a \in \mathit{Ch}$. Send and receive
transitions are paired up in system transitions:
\begin{equation*}
  \infer[\textsf{binary-comm}]
  {(g, \bar l) \to (g'', \bar l[p_1/l'_1][p_2/l'_2])}
  {(g, \bar l[p_1]) \stackrel{p_1,\, a!}{\longrightarrow} (g', l'_1)\quad
   (g', \bar l[p_2]) \stackrel{p_2,\, a?}{\longrightarrow} (g'', l'_2)\quad
    p_1 \not= p_2\quad a \in \mathit{Ch}}
\end{equation*}
Note that the effect of the send transition (on global state) occurs
prior to the execution of the receive transition; this means that
transfer of data can easily be realised with the help of additional
global variables.

\paragraph{Horn constraints.}
Recall that Fig.~\ref{fig:homogeneousHorn} contains two clauses,
\eqref{eq:homo2} and \eqref{eq:homo2b}, that model local process
transitions. Since communication through a channel implies that two
process transitions take place simultaneously (say, for processes
$p_s, p_r \in P$), it is now necessary to distinguish four cases (and
add four clauses) to characterise how a $k$-indexed invariant about
processes~$Q_k = \{p_1, \ldots, p_k\} \subseteq P$ is affected:
clause~\eqref{eq:binComm1} for the case $\{p_s, p_r\} \subseteq Q_k$
(this case disappears for $k = 1$); clause~\eqref{eq:binComm2} for
$p_s \in Q_k$, but $p_r \not\in Q_k$; clause~\eqref{eq:binComm3} for
$p_r \in Q_k$, but $p_s \not\in Q_k$; and clause~\eqref{eq:binComm4}
for $p_s, p_r \not\in Q_k$.  The clauses are instantiated for every
channel~$a \in \mathit{Ch}$:

\vspace*{-1ex}
{\small
\begin{align}
  \label{eq:binComm1}
    R(\q g'', \q p_1, \q l'_1, \q p_2, \q l'_2, \ldots, \q p_k, \q l_k)
    ~~\leftarrow~~
  &
  \begin{array}{@{}l@{}}
    \distinct(\q p_1, \ldots, \q p_k) \wedge
    \smash{\big((\q g, \q l_1) \stackrel{\q p_1,\, a!}{\longrightarrow}
      (\q g', \q l'_1)\big)}
    \wedge
    \smash{\big((\q g', \q l_2) \stackrel{\q p_2,\, a?}{\longrightarrow}
      (\q g'', \q l'_2)\big)}\wedge\mbox{}
    \\
    R(\q g, \q p_1, \q l_1, \q p_2, \q l_2, \ldots, \q p_k, \q l_k)
  \end{array}
  \\
  \label{eq:binComm2}
    R(\q g'', \q p_1, \q l'_1, \q p_2, \q l_2, \ldots, \q p_k, \q l_k)
    ~~\leftarrow~~
  &
  \begin{array}{@{}l@{}}
    \distinct(\q p_0, \ldots, \q p_k) \wedge
    \smash{\big((\q g, \q l_1) \stackrel{\q p_1,\, a!}{\longrightarrow}
      (\q g', \q l'_1)\big)}
    \wedge
    \smash{\big((\q g', \q l_0) \stackrel{\q p_0,\, a?}{\longrightarrow}
      (\q g'', \q l'_0)\big)}\wedge\mbox{}
    \\
    \mathit{RConj}(0, \ldots, k)
  \end{array}
  \\
  \label{eq:binComm3}
    R(\q g'', \q p_1, \q l'_1, \q p_2, \q l_2, \ldots, \q p_k, \q l_k)
    ~~\leftarrow~~
  &
  \begin{array}{@{}l@{}}
    \distinct(\q p_0, \ldots, \q p_k) \wedge
    \smash{\big((\q g, \q l_0) \stackrel{\q p_0,\, a!}{\longrightarrow}
      (\q g', \q l'_0)\big)}
    \wedge
    \smash{\big((\q g', \q l_1) \stackrel{\q p_1,\, a?}{\longrightarrow}
      (\q g'', \q l'_1)\big)}\wedge\mbox{}
    \\
    \mathit{RConj}(0, \ldots, k)
  \end{array}
  \\
  \label{eq:binComm4}
    R(\q g'', \q p_3, \q l_3, \q p_4, \q l_4, \ldots, \q p_{k+2}, \q l_{k+2})
    ~~\leftarrow~~
  &
  \begin{array}{@{}l@{}}
    \distinct(\q p_1, \ldots, \q p_{k+2}) \wedge
    \smash{\big((\q g, \q l_1) \stackrel{\q p_1,\, a!}{\longrightarrow}
      (\q g', \q l'_1)\big)}
    \wedge
    \smash{\big((\q g', \q l_2) \stackrel{\q p_2,\, a?}{\longrightarrow}
      (\q g'', \q l'_2)\big)}\wedge\mbox{}
    \\
    \mathit{RConj}(1, \ldots, k+2)
  \end{array}
\end{align}

}

\subsection{Unbounded Barrier Synchronisation}
\label{sec:barriers}

Besides rendezvous between pairs of processes, also barrier
synchronisation involving an unbounded number of processes can be
represented naturally in our model. Barriers turn out to be a powerful
primitive to encode other forms of communication, among others
Uppaal-style broadcast channels and BIP-style interactions
(Sect.~\ref{sec:bip}); of course, barriers are also highly relevant
for analysing concurrent software programs.  We assume a finite
set~$\mathit{Ba}$ of barriers, and denote process transitions
synchronising at barrier~$b \in \mathit{Ba}$ by $\smash{(g, l)
  \stackrel{p,\, b}{\longrightarrow} (g', l')}$. For simplicity, we
require \emph{all} processes in a system to participate in every
barrier synchronisation; a more fine-grained definition of the scope
of a barrier can be achieved by adding neutral transitions $\smash{(g,
  l) \stackrel{p,\, b}{\longrightarrow} (g, l)}$ to those processes
that are not supposed to be affected by $b$.

Barriers give rise to the following system transition:
\begin{equation*}
  \infer[\textsf{barrier}]
  {(g, \bar l) \to (g, \bar l')}
  {\big\{\;
    (g, \bar l[p]) \stackrel{p,\, b}{\longrightarrow} (g'_p, \bar l'[p])
    \;\big\}_{p \in P}
    \qquad
    b \in \mathit{Ba}}
\end{equation*}
Note that the system transition does not modify global state, but
alters all local state components simultaneously; the motivation for
this definition is to avoid clashes resulting from an
unbounded number of global state updates.

\paragraph{Horn constraints.}
Barrier synchronisation can be represented by a simple Horn constraint
(instantiated for every barrier~$b \in \mathit{Ba}$) stating that all
processes considered by a $k$-indexed invariant can do a transition
simultaneously:

\vspace*{-1ex}
{\small
\begin{equation}
    R(\q g, \q p_1, \q l'_1, \q p_2, \q l'_2, \ldots, \q p_k, \q l'_k)
    ~~\leftarrow~~
  \begin{array}{@{}l@{}}
    \distinct(\q p_1, \ldots, \q p_k) \wedge
    \smash{\big((\q g, \q l_1) \stackrel{\q p_1, b}{\longrightarrow}
    (\q g'_1, \q l'_1)\big) \wedge \cdots \wedge
    \big((\q g, \q l_k) \stackrel{\q p_k, b}{\longrightarrow}
    (\q g'_k, \q l'_k)\big)} \wedge\mbox{}
    \\
    R(\q g, \q p_1, \q l_1, \q p_2, \q l_2, \ldots, \q p_k, \q l_k)
  \end{array}
\end{equation}

}

\subsection{BIP Interactions}
\label{sec:bip}

BIP (Behaviour, Interaction, Priority)~\cite{DBLP:conf/atva/BensalemBSN08} is a
framework for designing component-based systems.  The BIP model of a
component consists of an interface (a set of ports) and a behaviour (an
automaton with transitions labelled by ports).  Components are composed
by a set of connectors that determine the interaction pattern among
the components. In general, when several interactions are possible the
system chooses the one which is maximal according to some given strict
partial order (priority). For sake of presentation, we concentrate on
a special case of interactions, \emph{rendezvous,} for which
priorities are irrelevant; the interactions in
Fig.~\ref{fig::bip-example} are all in the form of
rendezvous. However, other forms of interaction
provided by BIP (including interaction governed by priorities, and
ports that act as triggers) can be handled in our framework as well.

To define BIP rendezvous, we assume that $\mathit{Port}$ is a finite
set of ports, and $I \subseteq \cal{P}(\mathit{Port})$ is a set of
\emph{interactions.} A transition of process~$p \in P$ interacting
through port~$a \in \mathit{Port}$ is denoted by $\smash{(g, l)
  \stackrel{p,\, a}{\longrightarrow} (g', l')}$. The system transition
for an interaction~$\{a_1, \ldots, a_m\} \in I$ (with $m$ distinct
ports) is defined by the following rule; as a premise of the rule, it
is required that all processes of the system arrived at a point where
local (non-interacting) transitions are disabled
(\raisebox{0ex}[2.5ex]{$(g, \bar l[p]) \stackrel{p}{\not\to}$}), but
$m$ distinct processes~$p_1, \ldots, p_m$ are available that offer
interaction through ports~$a_1, \ldots, a_m$, respectively:
\begin{equation*}
  \infer[\textsf{bip-comm}]
  {(g, \bar l) \to (g, \bar l[p_1/l'_1]\cdots[p_m/l'_m])}
  {
      \big\{\;
      (g, \bar l[p_j]) \stackrel{p_j,\, a_j}{\longrightarrow} (g'_j, l'_j)
      \;\big\}_{j = 1,\ldots,m}
      & \{a_1, \ldots, a_m\} \in I
      &
      \big\{\;
      (g, \bar l[p]) \stackrel{p}{\not\to}
      \;\big\}_{p \in P}
      &
      \distinct(p_1, \ldots,  p_m)
  }
\end{equation*}

For the purpose of analysis within our framework, we reduce BIP
rendezvous to barrier synchronisation as in
Sect.~\ref{sec:barriers}. We make two simplifying assumptions:
\begin{inparaenum}
\item in no state~$(g, l)$ of a process~$p \in P$ both local
  transitions ($(g, l) \stackrel{p}{\to} \cdots$) and interacting
  transitions ($\smash{(g, l) \stackrel{p,\, a}{\longrightarrow}
    \cdots}$) are enabled; this will ensure the third premise
  of \textsf{bip-comm}; and
\item no two processes~$p_1, p_2 \in P$ share the same port~$a \in
  \mathit{Port}$.
\end{inparaenum}
Both assumptions can be established through suitable transformations
of a system.

We then encode BIP interaction using a single barrier~$\{b\} =
\mathit{Ba}$. To distinguish interactions, we choose a bijection~$h :
I \to \{1, \ldots, |I|\}$ that provides a unique integer as label for
each interaction, and add a global variable~$\mathit{iact}$ (part of
the global state~$g \in G$) ranging over $\{1, \ldots, |I|\}$.  In
addition, we denote the ports used by a process~$p \in P$ by
$\mathit{Port}_p \subseteq \mathit{Port}$. Each process~$p \in P$ of
the system is modified as follows:
\begin{itemize}
\item each interacting transition $\smash{(g, l) \stackrel{p,\,
      a}{\longrightarrow} (g', l')}$ is replaced with two barrier
  transitions. The first barrier transition is $\smash{(g, l)
    \stackrel{p,\, b}{\longrightarrow} (g', l')}$, and guarded with
  the test $a \in h^{-1}(\mathit{iact})$. The second transition is
  $\smash{(g, l) \stackrel{p,\, b}{\longrightarrow} (g, l)}$, i.e.,
  does not cause state changes, and is guarded with $\mathit{Port}_p
  \cap h^{-1}(\mathit{iact}) = \emptyset$.
\item a transition~$(g, l) \stackrel{p}{\to} (g[\mathit{iact}/*], l)$
  non-deterministically assigning a value to the variable
  $\mathit{iact}$ is added to the process~$p$; this transition is
  always enabled.
\end{itemize}



\section{Experimental Evaluation}
\label{sec:experiment}

\begin{figure}[tb]
 \begin{scriptsize}
 \begin{tabular*}{\textwidth}{@{\extracolsep{\fill}} l|l|l|l|p{2cm}|l}
  \textit{Benchmark}  & $\sharp Cl_{i}$ &  $\sharp Cl_{f}$ & $N^{th}$ \textit{(sec)} & \textit{Inv Schema}& \textit{Total (sec)}\\
  \hline
  Temperature Control System (unsafe) & 48 & 110 & 0.37 & (1,1,1) & 3.86
  \\[0.5ex]
  Temperature Control System & 48 & 110 & 1.12 & (1,1,1)  & 4.31
  \\[0.5ex]
  Fischer & 47 & 221 & 5.62  & (2,1)  & 12.21
  \\[0.5ex]
  Fischer (unsafe) & 47 & 221 & 2.84 & (2,1) & 8.91
  \\[0.5ex]
  CSMA/CD & 50 & 162 & 3.60 &  (2,1) & 8.22
  \\[0.5ex]
  CSMA/CD (unsafe) & 50 & 793 & 2.91 & (4,1) & 13.81
  \\[0.5ex]
  Lynch-Shavit & 50 & 299 & 66.11 & (2) & 70.10
  \\[0.5ex]
  Lynch-Shavit (unsafe) & 50 & 299 & 2.58 & (2) & 5.35
  \\[0.5ex]
  Train Crossing &28  & 686 & 2.51 & (1,3) & 8.84
  \\[0.5ex]
  Train Crossing (unsafe) & 28 & 240 & 1.53 & (1,2) & 3.91
  \\[0.5ex]
 \end{tabular*}

 \end{scriptsize}
\caption{Runtime for verifying the benchmarks. Experiments were done on an Intel Core i7 Duo 2.9 GHz with 8GB of 
RAM. Columns $\sharp Cl_{i}$ and $\sharp Cl_{f}$ indicate the number of clauses required to model the corresponding benchmark for 
the initial iteration and final iteration of the
counterexample-guided refinement of invariant schemata (Sect.~\ref{section:cegris}), respectively. The 
$N^{th}$ column indicates the time required to verify the benchmark on the final iteration. The 
\textit{Inv Schema} column contains the invariant schema required for the final (successful) iteration and \textit{Total} is the 
full verification time required.\label{fig:table}}
\end{figure} 

\enlargethispage{1ex}
We have integrated our technique into the predicate abstraction-based
model checker Eldarica~\cite{eldarica-tool}. 
In Table~\ref{fig:table} we show results
for benchmarks\footnote{Available at:
  \url{http://lara.epfl.ch/w/horn-parametric-benchmarks}} encoding timed models, verifying
natural safety properties of the models. All models but the temperature
control system (Sect.~\ref{sec:bipExample}) are unbounded; finite
instances of which are commonly used as benchmarks for model checkers.
Most of the benchmarks were originally specified as Uppaal timed
automata. For each benchmark we provide a correct version and
an unsafe version, to demonstrate the ability of our tool to prove
correctness and provide counter-examples for incorrect benchmarks. The
Fischer benchmarks contain an observer process, which is
the second component of the invariant schema.
The results demonstrate the feasibility of our approach. Further, it
can be seen that the majority of the benchmarks require stronger
invariant schemata, highlighting the benefits of $k$-indexed
invariants. Most benchmarks are solved within a few seconds.







\end{document}